\documentclass[APS,Times2COL]{WileyNJDv5} 
\usepackage[utf8]{inputenc}
\usepackage{textcomp}
\usepackage{newunicodechar}
\newunicodechar{−}{\ensuremath{-}}

\newcommand{\Sabs}[1]{\left\lceil #1 \right\rfloor}

\newcommand{\abs}[1]{\left| #1 \right|}
\newcommand{\norm}[1]{\left|\left| #1 \right|\right|}

\usepackage{xfrac}

\usepackage{siunitx}

\articletype{Research Article}%

\received{Date Month Year}
\revised{Date Month Year}
\accepted{Date Month Year}
\journal{Journal}
\volume{00}
\copyyear{2023}
\startpage{1}

\begin{document}

\title{Slip, Differentiate, Observe: State and Parameter Estimation for Rate and State Friction from Noisy Data}

\author[]{David Michael Riley }

\author[]{Diego Guti\'errez-Oribio}

\author[]{Ioannis Stefanou}

\authormark{Guti\'errez-Oribio \textsc{et al.}}
\titlemark{PLEASE INSERT YOUR ARTICLE TITLE HERE}

\address[]{ \orgname{IMSIA, UMR 9219, CNRS, EDF, ENSTA Paris, Institut Polytechnique de Paris},\country{France}}

\corres{Corresponding author David Riley. \email{david.riley@ensta.fr}}



\abstract[Abstract]{Quantifying frictional properties of interfaces remains a major challenge in both terrestrial and extraterrestrial geomechanics, where available samples, laboratory apparatuses, and geophysical observations are inherently limited. We introduce an analytic and numerical framework, grounded in nonlinear control theory, to infer the emergent frictional behavior of seismic faults. From noisy slip measurements, we first reconstruct the slip rate and frictional response in finite time using a Robust Exact Filtering Differentiator (REFD) that attenuates measurement noise. Building on these reconstructions, we design an exponentially convergent adaptive-gain observer that estimates the internal state variable and the key parameters ($a−b$) and $d_c$ of the rate-and-state friction (RSF) law, widely used in fault mechanics. Numerical experiments show that, in fast-slip regimes where data are sufficiently rich, the method recovers RSF parameters with errors on the order of 20 \% and accurately tracks the RSF state variable despite noise contamination, whereas slowly varying sliding periods lack the observability required for reliable estimation. We also establish observability and identifiability conditions for the extended system, enabling the determination of additional parameters and outlining pathways for more advanced control-theoretic approaches to friction and state identification in fault systems. Although we apply the approach to a reduced spring–slider analogue, it improves on classical RSF calibration methods that depend on laboratory access to shear and normal stress. It also offers convergence guarantees and explicit error bounds, and it can further support model-based inversions that embed RSF in forward simulations.}



\maketitle

\renewcommand\thefootnote{}
\footnotetext{\textbf{Abbreviations:} RSF, Rate and State Friction; REFD, Robust Exact Filtering Differentiator}

\renewcommand\thefootnote{\fnsymbol{footnote}}
\setcounter{footnote}{1}

\section{Introduction}
Frictional faults are central to a wide range of geomechanical phenomena, spanning from tectonic earthquakes and slow slip on plate-boundary faults~\cite{marone1998laboratory,obara2016connecting} to induced seismicity in reservoirs~\cite{ellsworth2013injection,ge2022induced} and fault reactivation in engineered underground structures~\cite{rutqvist2013modeling}. Similar frictional interfaces are also inferred on other planetary bodies, including lunar lobate scarps~\cite{williams2013fault,watters2015global} and Martian thrust faults~\cite{klimczak2018topographic}, where direct access to the fault zone is impossible, and data comes only from remote sensing and sparse geophysical observations. In all of these settings, the evolution of shear resistance along a fault controls whether slip proceeds steadily or in abrupt, seismic events. Laboratory experiments on fault gouges and analog materials have shown that some interfaces are velocity-strengthening, with steady-state friction increasing with slip rate, whereas others are velocity-weakening, with steady-state friction decreasing as slip rate increases~\cite{dieterich1979modeling,mair1999friction,ikari2021velocity,stathas2023fault,masi2025hydrodynamics}. A widely used empirical framework for describing such behaviour is Rate and State Friction (RSF), in which the friction coefficient depends on both slip rate and an internal state variable~\cite{dieterich1979modeling,ruina1983slip}. Depending on the parameter set, RSF can reproduce either velocity-strengthening or velocity-weakening responses and generate periodic frictional instabilities reminiscent of earthquakes~\cite{10.1115/1.3167042,gu1984slip,ciardo2025nonlinear}, and it has therefore become a standard ingredient in models for both terrestrial and extraterrestrial faulting.  In its simplest form, the RSF law is characterised by a small set of calibration parameters, typically denoted $a$ $b$, and $d_c$, which control the direct velocity effect, the evolution of state effect, and a characteristic slip distance. In what follows, we focus on estimating a subset of these parameters, along with the internal state, from indirect observations.

The traditional calibration procedure for RSF is to apply velocity-step and slide-hold-slide experiments, then regress the steady and transient responses to estimate the relevant parameters~\cite{ruina1983slip,marone1998laboratory,carpenter2016influence}. To side-step the necessity of multiple experiments, Scuderi et al.~\cite{tinti2016evolution} performed stick-slip experiments with a double direct shear apparatus while constantly sending p-waves and measuring shear stress to obtain the RSF parameters. While successful, these protocols typically assume direct access to shear/normal loads and carefully designed loading histories. However, in many natural and extraterrestrial settings, neither shear traction nor the internal state can be measured directly, and even slip must be inferred indirectly from surface deformation, microseismicity, or geomorphic offsets, making it essential to develop methods that can reconstruct frictional state and parameters from limited observations. This has motivated a complementary class of approaches that infer frictional properties indirectly by embedding RSF in forward models and constraining them with sparse geodetic or seismic observations.

Within this class, a more advanced, fully model-based strategy infers frictional properties by embedding RSF into forward simulations of fault slip.  In adjoint-based four-dimensional variational assimilation (4D-Var), one prescribes a quasi-dynamic RSF model for a given fault geometry and elastic structure and then adjusts spatially distributed frictional parameters so that simulated surface displacements reproduce geodetic observations of afterslip or slow slip events~\cite{10.1093/gji/ggv289,kano2020adjoint}. Kano et al.~\cite{kano2020adjoint} showed that the misfit in such problems can be highly sensitive to the assumed observations and initial guess of friction parameters, and that jointly estimating initial conditions and friction parameters leads to pronounced non-uniqueness and trade-offs. Alternatively, the ensemble Kalman filter treats slip rate, the internal state, and selected RSF parameters as components of an augmented state vector and updates them sequentially from geodetic data within a fixed quasi-dynamic forward model for slow slip events~\cite{10.1093/gji/ggz415}. Using a steady-state rate-and-state approximation, Fukuda and Burbot~\cite{fukuda2025ensemble} employed ensemble Kalman inversion with Bayesian inference to estimate in spatially varying stress-normalized combinations  $(a-b)\bar{\sigma}$ and $\mu_0\bar{\sigma}$ alongside
spatially varying viscoelastic rheology parameters from synthetic postseismic GNSS data. They found that frictional parameters were well constrained only in fault regions experiencing large coseismic stress changes, while viscoelastic parameters could only be reliably estimated in mantle regions with strong viscous strains.

At the scale of individual earthquakes, physics-based dynamic and quasi-dynamic inversions have been used to estimate frictional parameters directly from seismo-geodetic observations. Wang~\cite{wang2018next} formulated a stress-driven friction model for the Parkfield section of the San Andreas fault and used interseismic, coseismic, and early postseismic GNSS data to invert for a depth- and space-dependent rate-strengthening parameter $(a-b)>0$ together with a small number of coefficients scaling effective normal stress from seismic velocity models. Schliwa et al.~\cite{schliwa2024linked} combined a 3-D dynamic rupture model with a quasi-dynamic afterslip model governed by a RSF law and performed a Bayesian dynamic source inversion for the 2004 M$_{\text{w}}$ 6 Parkfield earthquake, jointly fitting strong-motion and GPS data to estimate spatially variable prestress and several frictional parameters, while quantifying strong trade-offs between, for example, shear prestress and reference friction. More recently,~\cite{STIERNSTROM2024113447} used a dynamic-rupture model with RSF and an adjoint formulation to invert for spatially variable friction properties on a rough fault, but in practice restricted their inversions to a single parameter field at a time (either the direct-effect parameter $a(x)$ or the initial shear stress $\tau_0(x)$), explicitly to avoid the parameter trade-offs that arise in multi-parameter inversions. At the opposite end of model complexity, complementary Bayesian work in highly reduced spring–slider systems has shown that probabilistic calibration of $d_c$ from noisy synthetic time series is feasible~\cite{DANA2021171}.

In parallel, machine-learning approaches have begun incorporating RSF laws as soft constraints in neural network models. For a single spring-slider system,~\cite{https://doi.org/10.1029/2023JB027384} used a physics-informed neural network (PINN) to infer three frictional parameters $(a,\,a-b,\,d_c)$ from noisy slip-velocity time series over a complete slow-slip cycle. Extending this idea to subduction interfaces,~\cite{https://doi.org/10.1029/2024JB030256} employed two coupled PINNs to reconstruct spatially varying fields $a(x,y)$, $(a(x,y)-b(x,y))$, and $d_c(x,y)$, and in some cases the initial state, from either fault slip or GNSS data in a quasi-dynamic slow slip events model. Their results highlight the strong dependence of parameter recovery on observation geometry and on whether a given region is velocity-weakening or velocity-strengthening. Complementing these PINN-based approaches,~\cite{kaveh2025data} combined machine learning with data assimilation methods, using neural networks to learn reduced-order model dynamics from POD decompositions of earthquake sequences, then employing an Ensemble Kalman Filter for sequential state estimation from slip-rate observations. They raised important questions about the observability of the RSF state variable from slip-rate data alone, arguing, using chaos synchronization concepts, that slip-rate histories may contain sufficient information to reconstruct the full state, though without a formal observability analysis. At the laboratory scale, physics-informed deep learning has also been used to estimate RSF parameters directly from stick-slip experiments based on acoustic emission signals~\cite{borate2024physics}. While PINNs demonstrate strong estimation capabilities, they typically require large data sets and often exhibit robustness issues, as well as significant regularity requirements.

In this paper, we adopt a complementary approach based on nonlinear control and observer theory. Related ideas are widely used for estimating LuGre-type friction parameters and their internal state~\cite{gafvert1999friction,10.1115/1.1870036}, as well as nonsmooth Coulomb friction~\cite{friedland1992adaptive,odabacs2024novel}. These methods construct an observer, that is, a dynamical system that mirrors the governing equations, is driven by the same inputs as the physical system, and uses the mismatch between predicted and measured outputs to update estimates of unmeasured states and parameters. The structure of Rate and State Friction, however, differs substantially from the static and LuGre-type models treated in that literature, so existing schemes cannot be transferred directly. We therefore focus on a reduced-order analogue of a fault segment, the classical spring–slider model, assuming that slip (displacement) can be measured. First, we show that slip velocity and friction can be reconstructed from noisy slip records using a model-free and robust differentiator that filters noise and converges in finite time~\cite{b:Levant-Livne-2020}. Second, we design an exponentially convergent adaptive observer that estimates the internal state and the frictional parameters $a-b$ and $d_c$ from the reconstructed frictional signal. Numerical experiments illustrate the performance and limitations of the proposed scheme.

In the following Section~\ref{sec:case}, we introduce the reduced order spring-slider as well as the measurement signal of slip used in the forthcoming analysis. Following this, Section~\ref{sec:diff} introduces the Robust Exact Filtering Differentiator (REFD) and the resultant estimation of slip rate and friction. In Section~\ref{sec:psi} presents the observer for the state used in the Rate and State Friction model and the parameter estimation. Finally, Section~\ref{sec:discussion} provides a discussion on the observability of the RSF state and identifiability of its parameters.

For the purposes of this paper, we define the (set-valued) function $\lceil\cdot\rfloor^{\gamma}:=|\cdot|^{\gamma}\mathrm{sign}(\cdot)$, for any $\gamma\in \Re_{\geq 0}$ with $\mathrm{sign}(x)=\left\{\begin{array}{cc}
1 & x>0 \\ 
\left[-1,1\right] & x=0 \\ 
-1 & x<0
\end{array}   \right.$.

\section{Frictional Spring-slider}
\label{sec:case}

\begin{figure}[ht!]
\begin{center}
\includegraphics[width=\columnwidth,keepaspectratio]{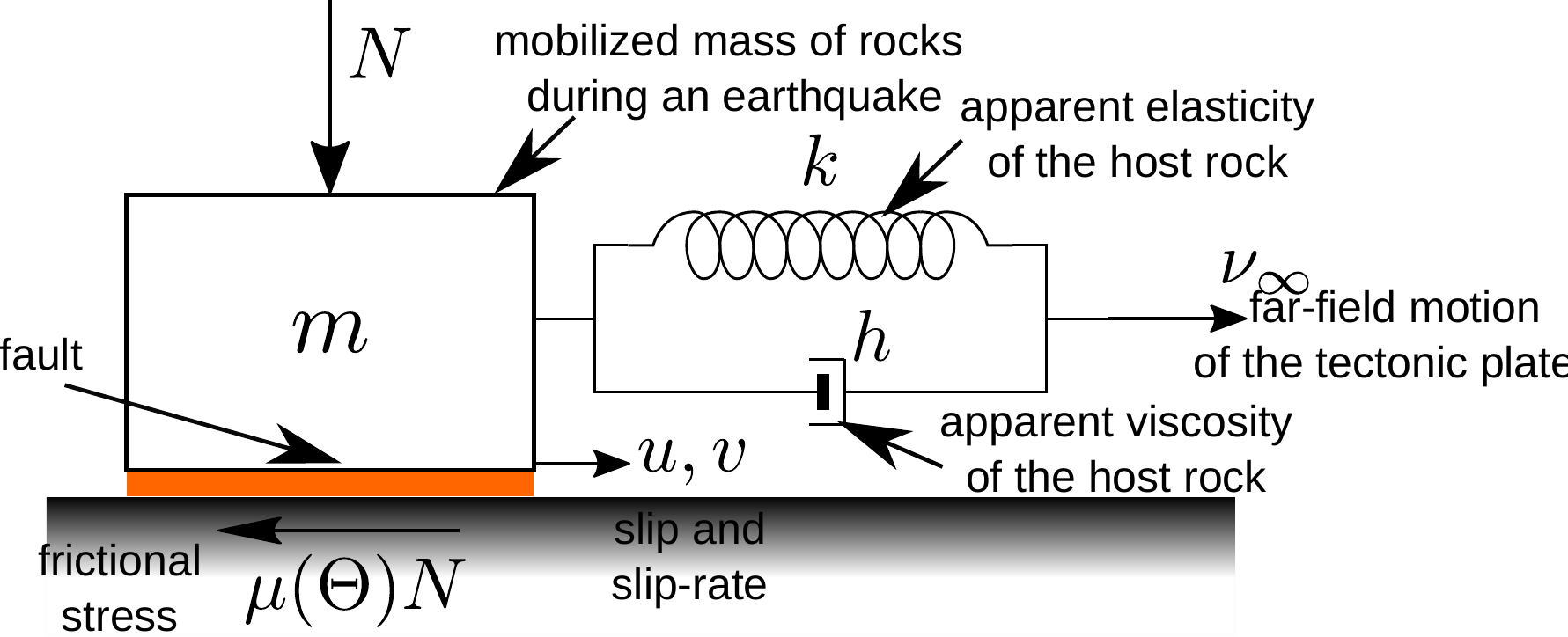}
\caption{Illustration of a one-degree-of-freedom spring slider model used as an analogue to a fault.  \label{fig1}}
\end{center}
\end{figure}

A widely used reduction for studying frictional interfaces is the one-degree-of-freedom spring--slider model. As illustrated in Figure~\ref{fig1}, a block of mass $m$ represents the mobilized portion of the crust during slip and is subject to a constant normal load $N$. The block is driven in shear by a loading point that moves at the far-field plate velocity $v_\infty$ and is connected to the block through a Kelvin--Voigt element consisting of a linear spring of stiffness $k$ in parallel with a dashpot of viscosity $h$. The generalized displacement of the block relative to the loading point is denoted by $u(t)$, and its slip rate (sliding velocity) is $v(t)$.

With this notation, the equations of motion read
\begin{align}
    \dot{u} &= v, 
    \label{eq:slip_kinematics} \\
    m \dot{v} &= -k\,(u - v_\infty t) - h\,(v - v_\infty) - \mu(\Theta)\,N,
    \label{eq:spring_slider_momentum}
\end{align}
where $\mu(\Theta)$ is the friction coefficient along the interface and $\Theta$ collects the variables on which friction may depend. In the RSF case considered later, $\Theta$ will include the slip rate and an internal state variable; however, for now, we keep $\mu$ in a generic form. Equation~\eqref{eq:slip_kinematics} simply states that $u(t)$ measures the inelastic slip of the block relative to the loading point, while~\eqref{eq:spring_slider_momentum} balances inertia with elastic, viscous, and frictional tractions. The material and rheological parameters used in this study are summarised in Table~\ref{tab:parameters}.

\begin{figure}[t]
\includegraphics[width=\columnwidth]{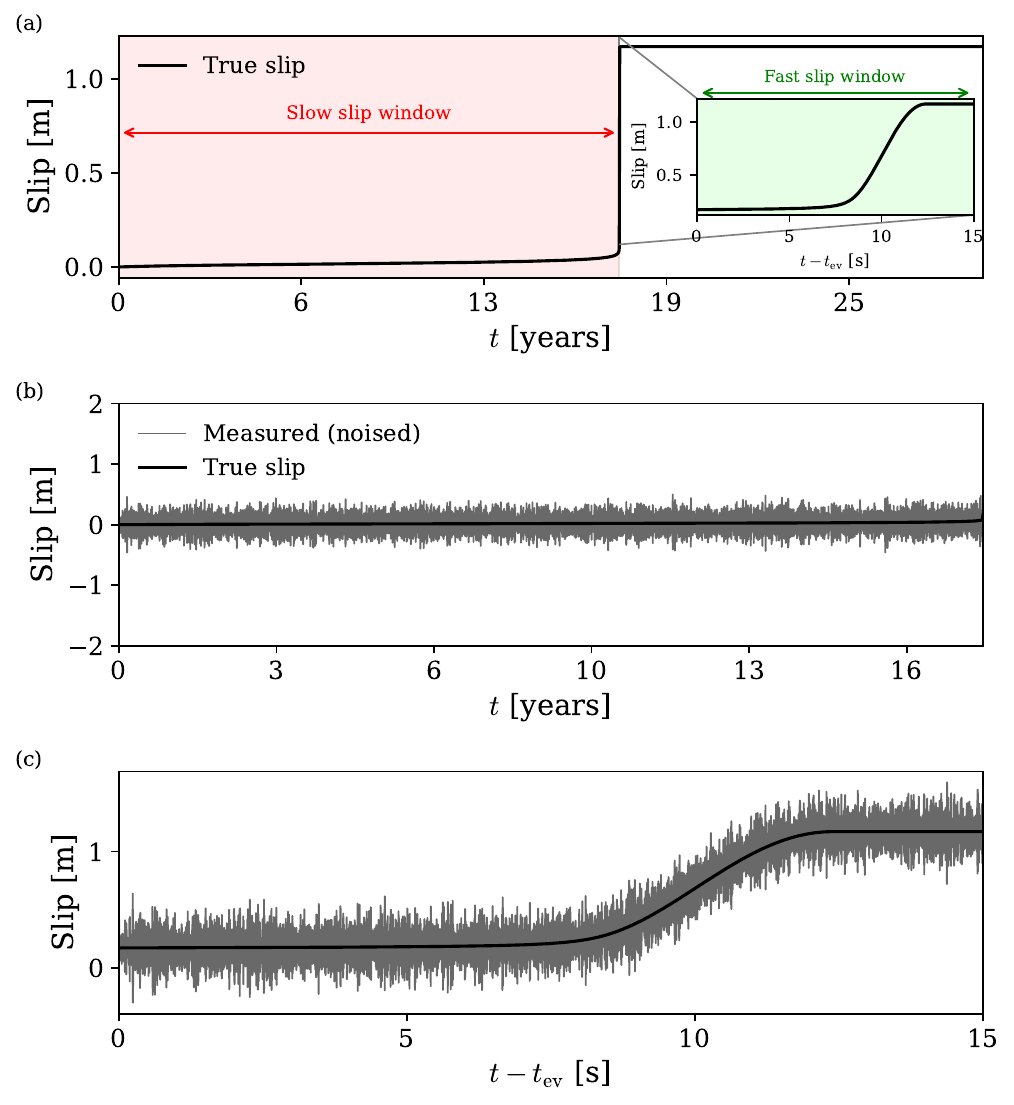}
\caption{True and noisy measurements of slip from spring-slider simulation. (a) True slip against time; the red shaded region marks the slow-slip interval, and the inset zooms into the fast-slip event. (b) Noisy measurement of slip during the slow-slip interval, plotted together with the true slip. (c) Noisy measurement of slip during the fast-slip interval, plotted together with the true slip. For clarity, the time axis in (c) and in the inset is plotted as $t-t_{ev}$ where $t_{ev}$ denotes the onset of the fast-slip event. \label{fig2}}
\end{figure}

Figure~\ref{fig2} (a) shows the true slip (i.e., noise-free) as a function of time; the red shaded region marks the slow-slip interval, and the inset zooms into the fast-slip event, shaded in green. During the slow-slip regime, we assume slip measurements are available every 6 hours, whereas during the fast-slip regime, we assume a sampling interval of 1 ms. This sharp contrast in sampling rates is intended to loosely mimic the difference between geodetic measurements of long-term deformation (hours to days between samples) and high-rate seismic or strong-motion recordings during rapid slip episodes.

In real applications, slip can be inferred from microseismicity and other geophysical measurements (\textit{e.g.}, GPS, LIDAR, and topographic surveys). To generate synthetic observations, we superimpose band-limited noise on the true slip. At each sampling time $t_i$ we define $y(t_i) = u(t_i) + \eta(t_i)$, where $\eta(t_i)$ is obtained by (i) drawing independent standard normal samples, (ii) passing them through a fourth-order zero-phase Butterworth bandpass filter with center frequency $f_{\text{c}} = 0.3/\Delta t$ (i.e., $0.6$ times the Nyquist frequency) and relative bandwidth $=0.3$, and (iii) rescaling the filtered sequence so that its root-mean-square (RMS) amplitude is a prescribed fraction of the RMS of the true slip on the corresponding interval. For the slow-slip window, we choose a large noise level, with $\mathrm{RMS}(\eta)\approx 10\,\mathrm{RMS}(u)$, to reflect the low signal-to-noise ratios typical of long-term surface measurements and inversions. For the fast-slip window, we use a more moderate noise level, with $\mathrm{RMS}(\eta)\approx 0.3\,\mathrm{RMS}(u)$, as a proxy to higher-quality measurements (e.g., well-instrumented strong-motion or laboratory data). In the theoretical analysis below, we assume only that the measurement error $\eta(t)$ is Lebesgue-measurable. Figure~\ref{fig2} (b) and (c) display the resulting noisy measurements $y(t)$ together with the underlying true slip $u(t)$ during the slow-slip and fast-slip intervals, respectively. The delineation between slow and fast slip was taken to be when $v>10^{-4}$ [m/s]~\cite{burgmann2018geophysics}. 

\section{Estimation of slip, slip-rate, and friction coefficient}
\label{sec:diff}

The measurements in Section~\ref{sec:case} are corrupted by noise, which, without proper filtering, would prevent spurious estimations of slip-rate and therefore friction. To obtain reliable estimates of the slip-rate $v$ and the friction coefficient $\mu$, we employ a Robust Exact Filtering Differentiator (REFD) \cite{b:Levant-Livne-2020}, given by
\begin{equation}
\begin{aligned}
  \dot{w}_1 &= -\lambda_6 L ^{\frac{1}{7}}\Sabs{w_1}^{\frac{6}{7}} + w_2, \\
  \dot{w}_2 &= -\lambda_5 L ^{\frac{2}{7}}\Sabs{w_1}^{\frac{5}{7}} + w_3, \\
  \dot{w}_3 &= -\lambda_4 L ^{\frac{3}{7}}\Sabs{w_1}^{\frac{4}{7}} + z_0 - y, \\
  \dot{z}_0 &= -\lambda_3 L ^{\frac{4}{7}}\Sabs{w_1}^{\frac{3}{7}} + z_1, \\
  \dot{z}_1 &= -\lambda_2 L^{\frac{5}{7}}\Sabs{w_1}^{\frac{2}{7}} + z_2, \\
  \dot{z}_2 &= -\lambda_1 L^{\frac{6}{7}}\Sabs{w_1}^{\frac{1}{7}} + z_3, \\
  \dot{z}_3 &= -\lambda_0 L \Sabs{w_1}^{0},
\end{aligned}
\label{eq:dif}
\end{equation}
where $y(t) = u(t) + \eta(t)$ denotes the noisy measurement of slip, and $z_0,z_1,z_2,z_3$ are the estimates of $u$, $\dot{u}=v$, $\ddot{u}=\dot{v}$, and $\dddot{u}=\ddot{v}$, respectively. The coefficients $\lambda_i>0$ and the gain $L>0$ are design parameters. In principle, estimating up to the second derivative of the slip is sufficient to reconstruct the system dynamics. However, the last channel of the REFD contains a discontinuous sign function. Thus, we estimate a third derivative of the slip to obtain a smoother approximation of the acceleration $\ddot{u}$.

Therefore, the REFD \eqref{eq:dif} provides a \emph{global}, \emph{finite-time} estimation of the time derivatives of the slip $u(t)$, while simultaneously filtering the measured signal through a third-order dynamics (the dynamics of $w_1(t),w_2(t),w_3(t)$), provided that $\abs{u^{(4)}(t)} \leq L$ and $\abs{\eta(t)} \leq \varepsilon$, for all time (see the proof in \cite{b:Levant-Livne-2020}). The estimation accuracy is given by
\begin{equation}
\begin{split}
  \abs{u^{(i)}(t) - z_i(t)} &\leq \gamma_i L \rho^{4-i}, 
  \quad i = 0,1,2,3, \\
  \rho &= \max \left\{\left(\frac{\varepsilon}{L}\right)^{\frac{1}{7}},\,\tau \right\},
\end{split}
\label{eq:dif_accu}
\end{equation}
for some constants $0<\gamma_i< \infty$ depending only on $\lambda_i$, and $\tau>0$ denoting the maximum sampling interval. Note that the global finite-time estimation of the slip and its derivatives becomes exact in the continuous-time, noise-free case ($\varepsilon = \tau = 0$).

Finally, we can estimate the friction term in \eqref{eq:spring_slider_momentum} with the estimates of slip, slip-rate, and acceleration from \eqref{eq:dif} as
\begin{equation}
    \hat{\mu}(t) =  -\frac{1}{N} \left[m z_2 + h\,(z_1 - v_\infty) + k\,(z_0 - v_\infty t) \right],
    \label{eq:mu_est}
\end{equation}
its precision can be obtained using \eqref{eq:dif_accu} and is given by
\begin{equation}
  \abs{\mu(t) - \hat{\mu}(t)} \leq \frac{L \rho^2}{N} \left[m \gamma_2 + h \gamma_1 \rho + k \gamma_0 \rho^2 \right].
    \label{eq:mu_accu}
\end{equation}

In reality, the slip measurements obtained are discrete signals as shown in Section~\ref{sec:case}. Therefore, the discrete implementation of the REFD for the results presented in this work was performed following \cite{b:https://doi.org/10.1002/rnc.70005}. Furthermore, the parameter $L$ was chosen as $L=10$ and the coefficients $\lambda_i$ were selected following \cite{b:7f0a759c989b4156997a1d79887dd0db} as $\lambda_6=7.09$, $\lambda_5=21.58$, $\lambda_4=36.46$, $\lambda_3=36.96$, $\lambda_2=22.48$, $\lambda_1=7.59$, $\lambda_0=1.1$. Additionally, the differentiator requires initial conditions for the slip: we set the slip to the noisy measured quantity, and the slip rate to the known far-field velocity $v_{\infty}$. The remaining derivatives were set equal to zero.

\begin{figure}[ht!]
\includegraphics[width=\columnwidth]{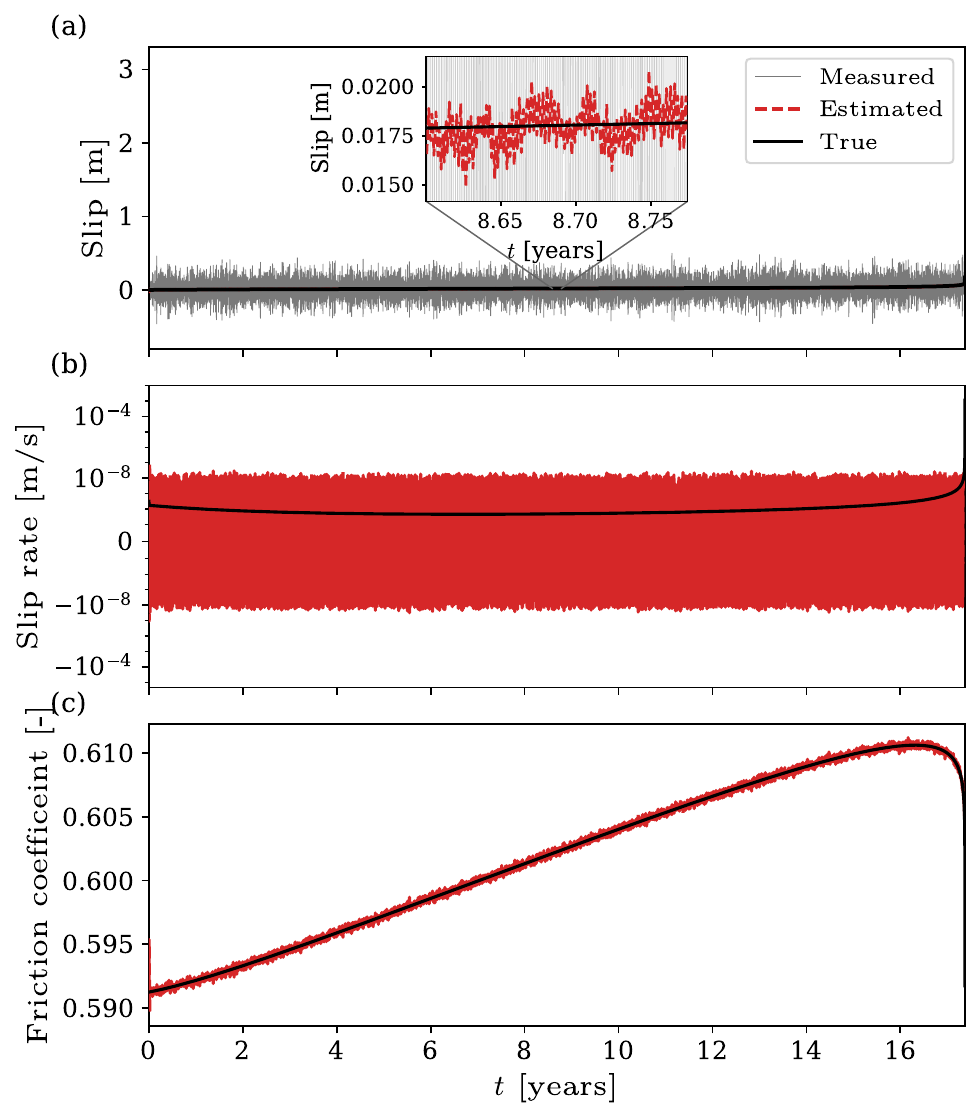}
\caption{Performance of the Robust Exact Filtering Differentiator (REFD) during the slow-slip regime. (a) Noisy slip measurements $y(t)$ (grey), true slip $u(t)$ (black), and REFD slip estimate (red dashed); 
the inset zooms into a short time interval and highlights the convergence of the estimate to the true signal. 
(b) True slip rate $v(t)$ (black) and REFD slip rate estimate (red) on a logarithmic scale. 
(c) True friction coefficient $\mu(t)$ (black) and reconstructed friction $\hat{\mu}(t)$. The subsequent fast-slip episode of the same signal is shown in Figure~\ref{fig4}.  \label{fig3}}
\end{figure}

\begin{figure}[ht!]
\includegraphics[width=\columnwidth]{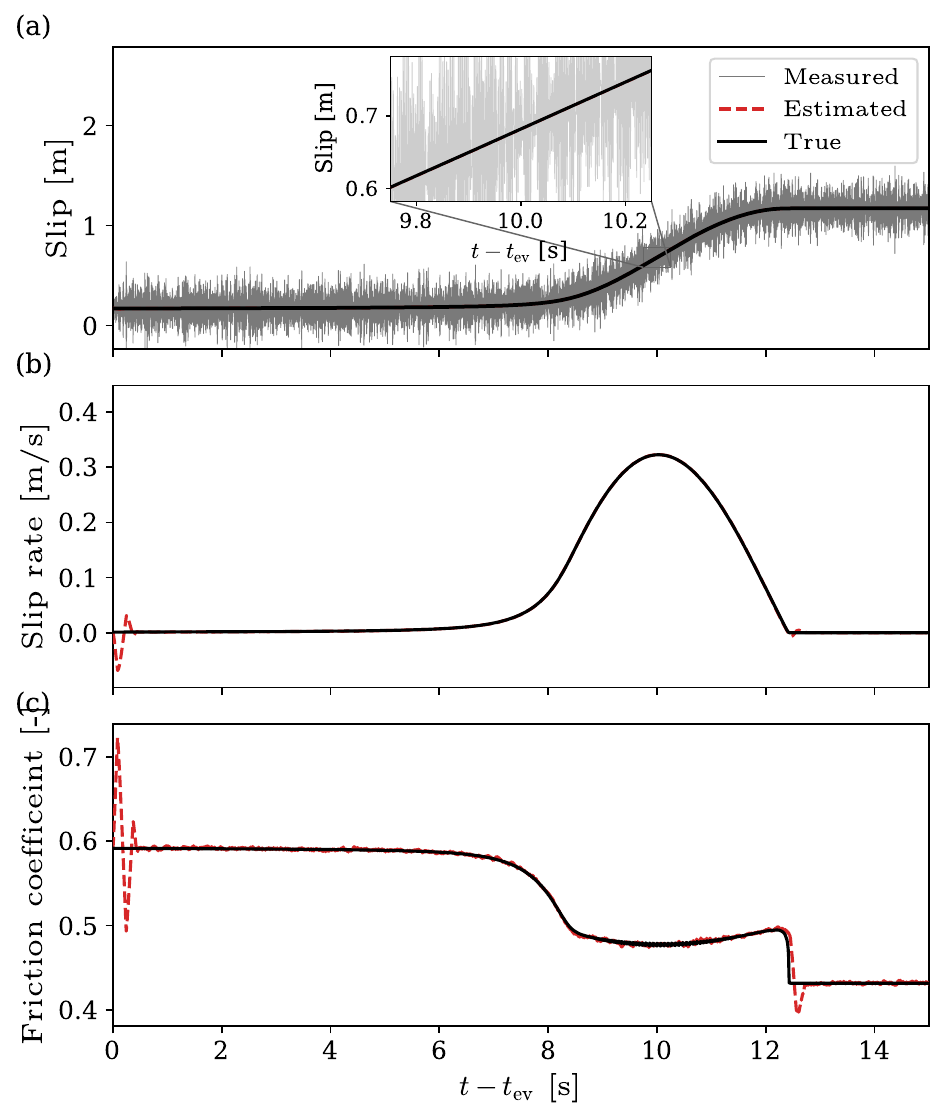}
\caption{Performance of the Robust Exact Filtering Differentiator (REFD) during the fast-slip regime. (a) Noisy slip measurements $y(t)$ (grey), true slip $u(t)$ (black), and REFD slip estimate (red dashed); 
the inset zooms into a short time interval and highlights the convergence of the estimate to the true signal. 
(b) True slip rate $v(t)$ (black) and REFD slip rate estimate (red) on a logarithmic scale. 
(c) True friction coefficient $\mu(t)$ (black) and reconstructed friction $\hat{\mu}(t)$.  \label{fig4}}
\end{figure}

In the slow-slip regime, Figure~\ref{fig3}~(a) shows that the measured signal is visually dominated by noise, so that the underlying slip $u(t)$ is essentially unrecognizable at the plotting scale. Nevertheless, the REFD output rapidly converges to the true slip and tracks its long-term evolution; the inset highlights the accuracy. The corresponding slip-rate estimates in panel~(b) are noisier, as expected for a differentiator operating on data with such a poor signal-to-noise ratio. However, the cloud of REFD estimates remains centered around the true velocity curve and resolves its slow temporal variation over several orders of magnitude in $v$. Most importantly for our purposes, the reconstructed friction coefficient $\hat{\mu}(t)$ in panel~(c), obtained from~\eqref{eq:mu_est}, is almost indistinguishable from the true friction $\mu(t)$ on the scale of the plot, apart from the deviation at the onset and termination of the interval. Figure~\ref{fig3} demonstrates that the REFD is able to extract the slowly varying component of the motion and deliver especially accurate friction estimates even when the raw slip data are extremely noisy. The continuation of the same signal through the fast-slip episode is examined in Figure~\ref{fig4}.

In the fast-slip regime, Figure~\ref{fig4} (a) shows that the noisy slip measurements still show significant high-frequency fluctuations. However, the estimated slip overlays the true slip curve throughout the dynamic event, and the inset again illustrates rapid convergence on the event timescale. The slip-rate estimates in Figure~\ref{fig4} (b) closely follow the true velocity peak, capturing both the rapid acceleration and deceleration phases. As a result, the reconstructed friction coefficient $\hat{\mu}(t)$ in Figure~\ref{fig4} (c) tracks the true friction through the rapid weakening associated with the fast-slip episode.

\section{Estimation of the frictional parameters and State}
\label{sec:psi}
In this section, we use the reconstructed friction $\hat{\mu}(t)$ and slip-rate estimate $z_1(t)$ from Section~\ref{sec:diff} to estimate the internal state $\psi$ and selected rate-and-state friction (RSF) parameters $a-b$ and $d_c$. To begin, express the general form of the state-evolution equation used in this work~\cite{https://doi.org/10.1002/zamm.201800263}
\begin{equation}
    \dot{\psi} = -A(\psi) + f(\abs{v}),
    \label{eq:state_evolution_general}
\end{equation}
where $A(\psi):\Re \to \Re$ and $f(\abs{v}):\Re_0^+ \to \Re$. Furthermore, we will assume the friction coefficient as a function of the state $\psi$ and velocity $v$, i.e., $\mu(\abs{v},\psi): \Re_0^+ \times \Re \to \Re_0^+$.

Some important assumptions on $\mu(\abs{v},\psi)$, $A(\psi)$ and $f(\abs{v})$ are state in the following \cite{https://doi.org/10.1002/zamm.201800263}:
\begin{assumption}\label{A1}
The function $\mu$ is \emph{nondecreasing} and continuous in its first argument.
\end{assumption}
\begin{assumption}\label{A2}
The function $\mu$ is uniformly Lipschitz in its second argument, i.e., we have
\begin{equation}
\abs{ \mu(r,\alpha)-\mu(r,\beta)} \le L_\mu \abs{ \alpha-\beta},
\end{equation}
$\text{for all }\alpha,\beta\in\Re \text{ and } r\ge 0$.
\end{assumption}
\begin{assumption}\label{A3}
The function $\mu$ can be bounded as follows:
\begin{equation}
0 \le \mu(r,\alpha) \le C_\mu\bigl(1+r+\abs{ \alpha}\bigr),
\end{equation}
$\text{for all }\alpha\in\Re \text{ and } r\ge 0$.
\end{assumption}
\begin{assumption}\label{A4}
The function $A$ is \emph{nondecreasing} and continuous. Furthermore, it is locally strongly monotone, i.e., there exists a subspace $\mathcal{K} \subset \Re$ such that:
\begin{equation}
(A(\alpha)-A(\beta))(\alpha-\beta) \geq L_\psi (\alpha-\beta)^2,
\end{equation}
$\text{for all }\alpha,\beta\in \mathcal{K}$.
\end{assumption}
\begin{assumption}\label{A5}
The function $f$ is Lipschitz, so that we have
\begin{equation}
\lvert f(r)-f(\nu)\rvert \le L_f\, \lvert r-\nu\rvert,
\qquad \text{for all } r,\nu\in\Re.
\end{equation}
\end{assumption}

For the application considered here, we adopt the following RSF law~\cite{https://doi.org/10.1002/zamm.201800263}
\begin{equation}
    \mu(\abs{v},\psi)
    = \mu_* 
    + a \,\ln\!\left(\frac{|v|}{v_*}\right)
    + b\psi,
    \label{eq:rsf_mu}
\end{equation}
where $\mu_*$ is a reference friction coefficient measured at a chosen reference slip rate $v_*$, and $a$ and $b$ are dimensionless empirical parameters. 

The evolution of law for the state variable is~\cite{Dieterich1979,Dieterich1981,https://doi.org/10.1002/zamm.201800263}
\begin{align}
    \dot{\psi} &= \frac{s}{d_c}, \quad s=v_* e^{- \psi} - \abs{v},
    \label{eq:state_evolution}
\end{align}
where $d_c$ is the characteristic slip distance. Equation~\eqref{eq:state_evolution} is the so-called aging law (also called the slowness law) for the state variable $\psi$. This evolution law fits the general structure of~\eqref{eq:state_evolution_general} with $A(\psi)=-\frac{v_* e^{- \psi}}{d_c}$ and $f(\abs{v})=-\frac{\abs{v}}{d_c}$. In this empirical law, $\psi$ is a state variable that captures time-dependent healing as it is related to the renewal of true contact asperities. Moreover, the original form of RSF is recovered by introducing $\theta=e^{\psi}$. 

The parameters used for all of the ground-truth simulations are presented in Table~\ref{tab:parameters}. In what follows, we treat the mechanical parameters $m,k,h,N,v_\infty$ as known quantities. This choice is motivated by the fact that, in both laboratory and field applications, the effective stiffness $k$ can usually be estimated from the shear modulus of the host rock and a characteristic fault dimension $L_x$, while $m$ and $N$ are determined from density, geometry, and depth-dependent overburden and pore-pressure, which are all relatively constrained quantities. The dashpot coefficient $h$ (or, equivalently, the damping ratio $\zeta$) is in practice less tightly constrained. We fix $\zeta=0.1$ as a representative weakly damped value and note that, for such low damping, viscous effects leave the natural period essentially unchanged and primarily act to smooth high-frequency oscillations, whereas the transition between stable/slow slip and stick–slip is governed mainly by the stiffness-to-critical-stiffness ratio and the RSF parameters $(a,b,d_c)$~\cite{tzortzopoulos2021controlling}. Thus, the value of $\zeta$ is not expected to modify the results found here.

\begin{table}[t]
\begin{center}
{\scriptsize
  \centering
  \caption{Model parameters used in the simulations.}
  \begin{tabular}{llcc}
    \hline
    Symbol & Description & Value & Unit \\
    \hline
    $L_x$        & Fault length                    & $5\times 10^{3}$       & m \\
    $G$          & Shear modulus                   & $3.0\times 10^{10}$    & Pa \\
    $N$   & Normal stress                   & $3.75\times 10^{7}$    & Pa \\
    $\rho$       & Density                         & $2.5\times 10^{3}$     & kg/m$^{3}$ \\
    $\zeta$      & Damping ratio                   & $0.1$                  & -- \\
    $m$    & Block mass    & $\rho L_x^3$     & kg \\
    $k$    & Apparent elasticity    & $G L_x$     & Pa $\cdot$ m \\
    $h$    & Dashpot viscosity    & $2 \zeta m \sqrt{k/ m}$     & Pa $\cdot$ m $\cdot$ s\\
    $v_\ast$     & Reference slip-rate             & $4\times 10^{-6}$      & m/s \\
    $v_\infty$   & Far field velocity               & $3.17 \times 10^{-10}$              & m/s \\
    $\mu_*$      & Static friction coefficient     & $0.55$                 & -- \\
    $a$ & RSF parameter $a$           & $1.0\times 10^{-2}$    & -- \\
    $b$ & RSF parameter $b$           & $1.5\times 10^{-2}$    & -- \\
    $d_c$        & Characteristic slip distance    & $10\times 10^{-3}$     & m \\
    \hline
  \end{tabular}
  \label{tab:parameters}}
\end{center}
\end{table}

With this structure in place, we now turn to the observer design. The goal is to use the friction estimate $\hat{\mu}(t)$ and the slip-rate estimate $z_1(t)$ obtained in Section~\ref{sec:diff} to reconstruct the internal state $\psi$ and the frictional parameters $a-b$ and $d_c$. To this end, we propose the following adaptive observer:
\begin{equation}
\begin{aligned}[c]
  \dot{\tilde{\mu}} &= \hat{b}\,\hat{\xi}\,\hat{s} + k_1 e,\\
  \dot{\hat{\psi}}  &= \hat{\xi}\,\hat{s} + k_2 e,\\
  \dot{\hat{b}}     &= k_3 e\,\hat{\xi}\,\hat{s},\\
  \dot{\hat{\xi}}&= k_4 e\,\hat{b}\,\hat{s},
\end{aligned}
\qquad
\begin{aligned}[c]
  e &= \hat{\mu} - \bar{a}\ln\bigl(\lvert z_1\rvert\bigr) - \tilde{\mu}, \\
  \hat{s} &= v_* e^{-\hat{\psi}} - \lvert z_1\rvert,
\end{aligned}
\label{eq:observer}
\end{equation}
where $\tilde{\mu}+\bar{a}\ln\bigl(\lvert z_1\rvert\bigr)$, $\hat{\psi}$, $\hat{b}$, and $\hat{\xi}$ denote the estimates of a $\hat{\mu}$, $\psi$, $b$, and $\xi=1/d_c$, respectively. The gains $k_1,k_2,k_3,k_4>0$ are design parameters, and $\bar{a}$ is a nominal guess of the friction parameter $a$.

Then, \emph{local exponential convergence} of the error vector 
$
x := \begin{bmatrix} e & \psi-\hat{\psi} & b-\hat{b} & \xi-\hat{\xi} \end{bmatrix}^{\!\top}
$
to the origin is ensured on the set
$
\mathcal{X}_r := \left\{ x\in\mathbb{R}^4 \,:\, \|x\|\le r,\ \psi,\hat{\psi}\in\mathcal{K}\right\},
$
provided the observer gains are chosen as
\begin{equation}
  k_1 > \frac{1 - S_{\max} r^2}{2}, 
  \quad 
  k_2 < 2 v_* L_\psi - S_{\max}, 
  \quad
  k_3>0, 
  \quad 
  k_4>0,
  \label{eq:gains_obs}
\end{equation}
where $\lvert s(t)\rvert \le S_{\max}$ for all $t \geq 0$. The steady-state estimation accuracy satisfies
\begin{equation}
  \|x\|
  \;\leq\;
  c \,\frac{L \rho^2}{h}
             \Bigl[m \gamma_2 + 2 h \gamma_1 \rho + k \gamma_0 \rho^2 \Bigr], 
  \label{eq:psi_accu}
\end{equation}
for some constant $0 < c < \infty$. In principle, this constant $c$ can be computed numerically for each specific configuration (see the proof in Appendix~\ref{app:proof}). Hence, bound~\eqref{eq:psi_accu} explicitly reflects the contributions of the differentiator and friction-reconstruction accuracies given in~\eqref{eq:dif_accu} and~\eqref{eq:mu_accu}.

To test the adaptive observer, it is crucial to operate in a regime of \emph{persistent excitation}. This ensures sufficient richness of the data over a finite period of time for parameter identifiability and guarantees convergence of classical adaptive laws (see, e.g., \cite{b:Ioannou-Sun,b:Narendra-Annaswamy}). Here, sufficient richness means that the regressor signals $\phi(t)$ (built from $v$, $\dot v$ and $\psi$) are \emph{persistently exciting}, i.e., over any window of length $T>0$ the matrix $\int_{t}^{t+T} \phi(\tau)\phi^\top(\tau)\,d\tau$ is uniformly positive definite, so that no direction in the parameter space is systematically unexcited. Such excitation is naturally provided only during the fast-slip regime. For this reason, we implement the adaptive observer exclusively in this regime. Moreover, we run it iteratively: after each fast-slip event, the final values of $\hat{b}$ and $\hat{\xi}$ are stored and used as initial conditions for the next event. This way, the observer accumulates information over successive slip episodes, allowing the parameter estimates $(\hat{b},\hat{\xi})$ to progressively converge to their true values, even though each fast-slip interval provides only a finite amount of excitation.

We assume a nominal guess $\bar{a} = 0.5\,a$ and initialize the parameter estimates as
$\hat{b} = 10 b$ and $\hat{\xi} = \tfrac{1}{10 d_c}$. Moreover, the initial conditions for the observer are $\tilde{\mu}(0) =\hat{\mu}(t_{ev})-\bar{a}\ln\bigl(\lvert z_1(t_{ev})\rvert\bigr)$ and $\hat{\psi}(0)=0$.  The observer gains are chosen as $k_1 = 5 \times 10^2$, $k_2 = 1 \times 10^3$, $k_3 = 10$, and $k_4 = 3 \times 10^6$. 

\begin{figure}[ht!]
\includegraphics[width=\columnwidth]{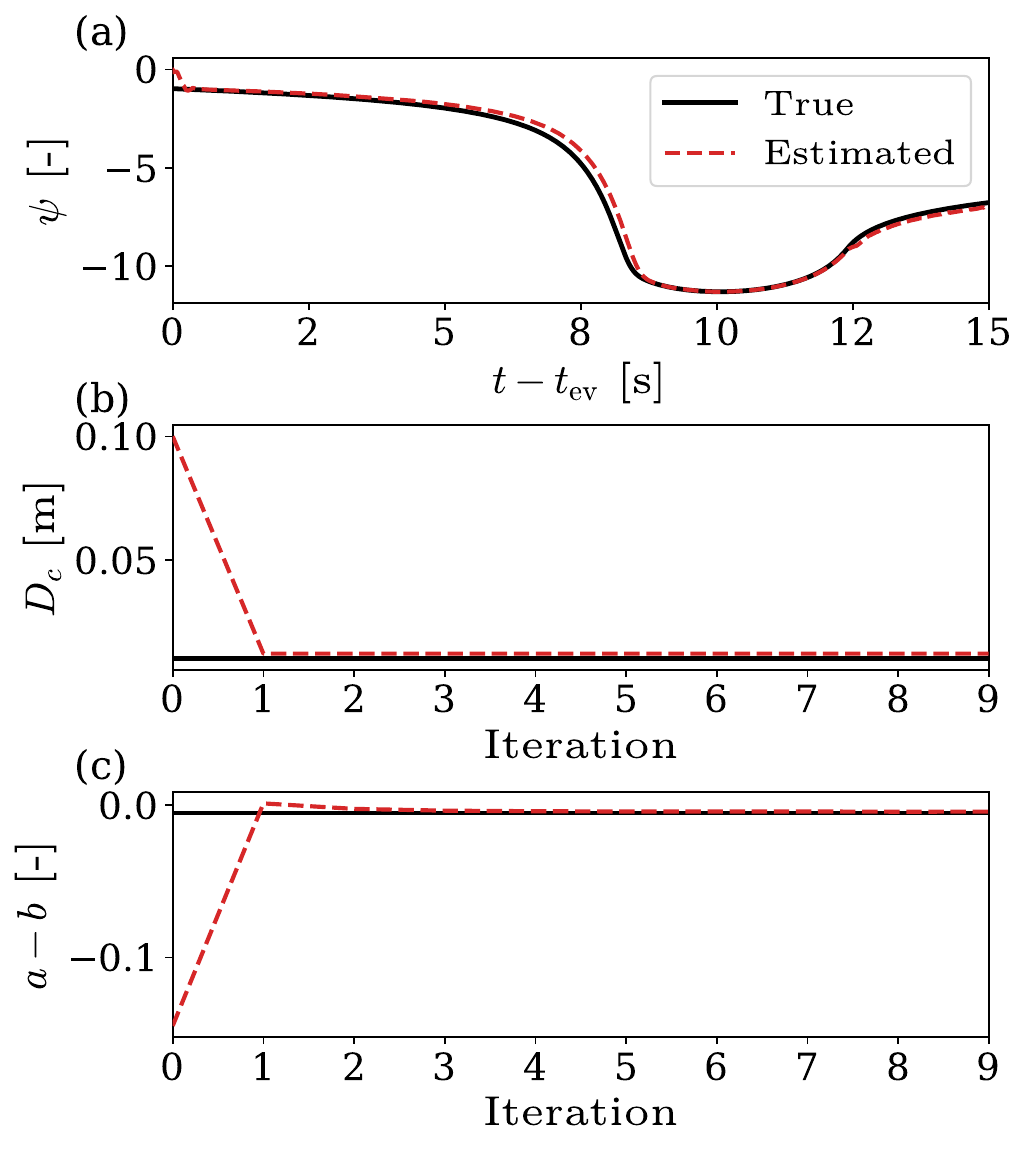}
\caption{Performance of the observer during the fast-slip regime. (a) Estimation for state $\hat{\psi}(t)$ (red dashed) and true state $\psi(t)$ (black) in time. 
(b) True $d_c$  (black) and estimated $\hat{d_c}$ against iterations. 
(c) True $a-b$  (black) and estimated $\hat{a}-\hat{b}$ against iterations.  \label{fig:observer_results}}
\end{figure}
The results in Figure~\ref{fig:observer_results} show that the adaptive observer reconstructs the internal state $\psi(t)$ with good accuracy. In addition, the parameter estimates converge to values close to the true ones, yielding a relative error of approximately $28.1\%$ for $d_c$ and $14.9\%$ for $a-b$. These results can be further reduced by selecting a nominal value $\bar{a}$ closer to the true parameter $a$.



\section{Discussion}
\label{sec:discussion}

One of the key features of this paper is the application of control-theoretic tools for state and parameter estimation of the spring-slider system with RSF. This gives a rigorous way to assess \emph{observability}, i.e., whether the internal states of a dynamical system can be reconstructed from its measured output, and \emph{identifiability}, which is whether it is possible to infer a dynamical system's parameters by measuring its output. In what follows, we append the RSF parameters to the state vector and assign them trivial dynamics (no time evolution). Thus, in this scenario, the observability of this augmented state from slip measurements becomes equivalent to recovering both the RSF state $\psi$ and the friction parameters, so observability and identifiability coincide in our setting.

For the spring-slider, we consider the augmented state
$
  z := \left[ \mu,\psi, b, \xi \right]^\top,
$
and we treat the slip-rate $v(t)$ and its derivative $\dot v(t)$ as known exogenous inputs, and the friction coefficient $\mu$ as the measured output. The dynamics can be written as
\begin{equation}
\begin{split}
    \dot{\mu} &= a \frac{\dot{v}}{v} + b\,\xi\, s, \\
    \dot{\psi} &= \xi\, s, \quad  s=v_* e^{- \psi} - \abs{v}, \\
    \dot{b} &= 0,\\ 
    \dot{\xi} &= 0,\\
    y &= \mu.
\end{split}
\label{eq:rsf_augmented}
\end{equation}
We write this compactly as $\dot z = f(z,w(t))$, $y=h(z)$, where
$w(t):=(v(t),\dot v(t))$ is the known input.

According to the Hermann-Krener rank condition \cite{b:HermannKrener1977}, the system
\eqref{eq:rsf_augmented} is locally observable at $z\in D\subset \Re^4$ if the codistribution generated by the differentials of the successive Lie derivatives of $h$ along $f$ has full dimension $4$. Equivalently, one can form the \emph{nonlinear observability matrix}
\begin{equation}
  \mathcal{O}(x)
  :=
  \left[
    h(x) ,
    L_f h(x) ,
    L_f^2 h(x) ,
    L_f^3 h(x)
  \right],
  \label{eq:O_def}
\end{equation}
and require $\operatorname{rank}\bigl(\mathcal{O}(x)\bigr) = 4$, $\forall z \in D$.

\begin{figure}[ht!]
\includegraphics[width=\columnwidth]{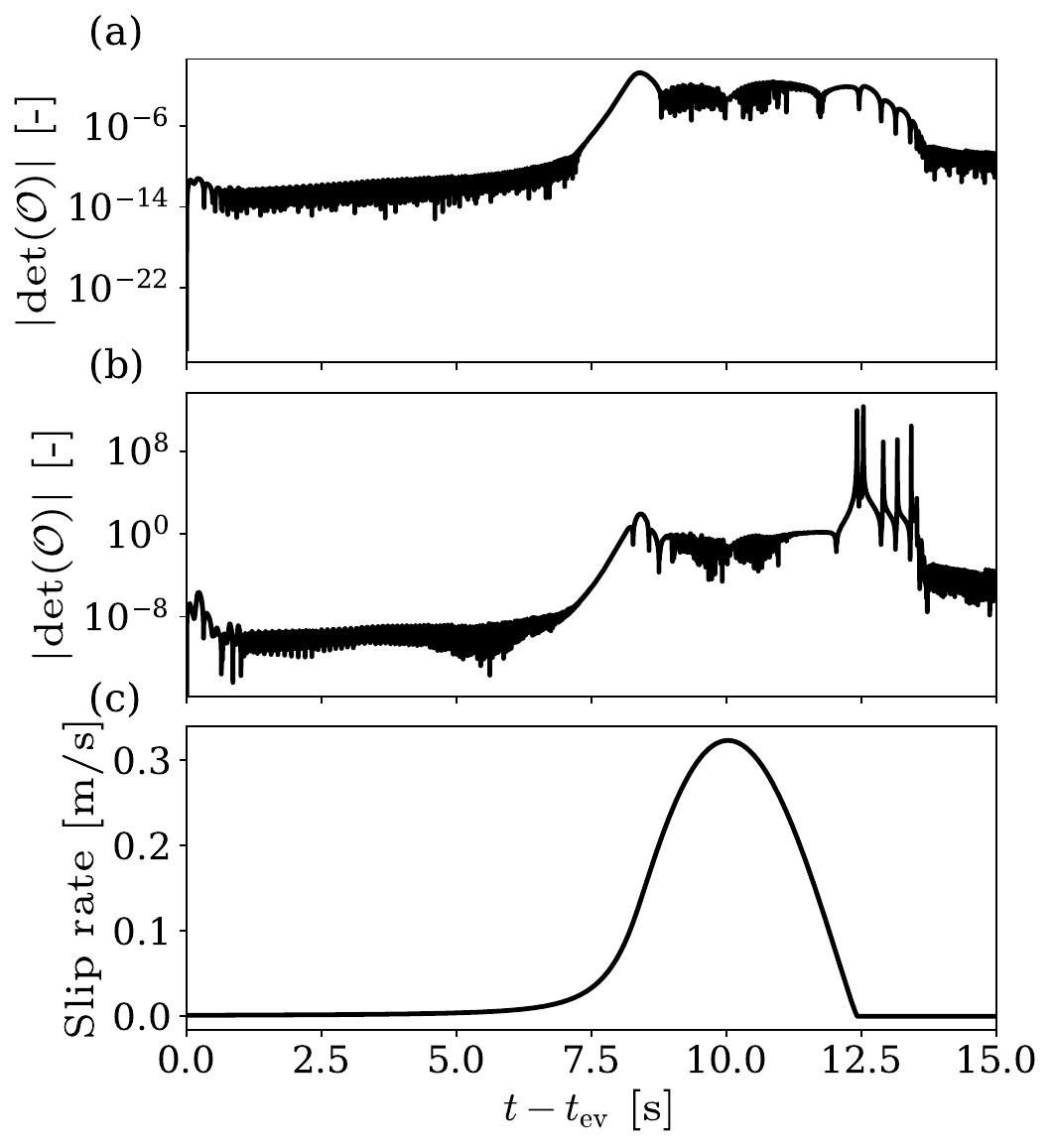}
\caption{Time evolution of observability. (a) Absolute value of the determinant of the observability matrix for the two-parameter frictional observability. 
(b) Same quantity for the three-parameter observability problem.
(c) Slip rate against shifted time where $t_{ev}$ is the time of the event. \label{fig:observability}}
\end{figure}

Due to the complexity of the observability matrix $\mathcal{O}(x)$, we evaluate it numerically along the simulated trajectories rather than in closed form. Figure~\ref{fig:observability} (a) shows the absolute value of the determinant $\lvert\det\mathcal{O}(t)\rvert$ for the two-parameter estimation problem ($b$ and $\xi = 1/d_c$), corresponding to system~\eqref{eq:rsf_augmented}. A value $\lvert\det\mathcal{O}(t)\rvert \approx 0$ indicates loss of local observability, since $\mathcal{O}(t)$ becomes close to rank-deficient. We observe that $\lvert\det\mathcal{O}(t)\rvert$ increases by several orders of magnitude during the dynamic phase of the event and remains largest when the slip rate is high (see Fig.~\ref{fig:observability} (c)), confirming that intervals of rapid slip carry the most information about the RSF parameters.

To investigate whether it is also possible to estimate the frictional parameter $a$, we augment system~\eqref{eq:rsf_augmented} with an additional constant parameter state $\dot{a}=0$ and recompute $\lvert\det\mathcal{O}(t)\rvert$ numerically. Figure~\ref{fig:observability} (b) shows that the largest peaks now
occur near the onset and termination of rapid slip ($t-t_{\text{ev}}\approx 7.5$~s and $t-t_{\text{ev}}\approx 12.5$~s), where $v(t)$ and $\psi(t)$ (see Figure~\ref{fig:observer_results}) vary most rapidly. This indicates that, at least from an observability perspective, an adaptive observer for the full three-parameter case $(a,b,d_c)$ is feasible and thus the system is identifiable. The actual design and Lyapunov analysis of such an observer, however, are nontrivial and
are left as future work.

For both the two- and three-parameter cases, $\lvert\det\mathcal{O}(t)\rvert$ remains essentially zero near the initial quasi-static state, indicating that the system is locally unobservable in that regime. Indeed, by injecting the steady-state regime into the observability matrix, we find that $\det(\mathcal{O})=0$, and therefore the system is not identifiable in this regime. 

\section{Conclusions}
\label{sec:conclusion}

In this work, we have developed a novel approach for estimating Rate and State Friction parameters from noisy slip measurements by combining robust differentiation with adaptive observer theory. Our method addresses a fundamental challenge in fault mechanics: inferring frictional properties when neither shear traction nor the internal state can be measured directly, as is typical in natural and extraterrestrial settings.

The key contributions of this study are threefold. First, we illustrated that the Robust Exact Filtering Differentiator (REFD) can accurately estimate slip velocity and friction coefficient from extremely noisy slip measurements, even when the signal-to-noise ratio is as low as 0.1 during slow-slip periods. This robustness is crucial for applications where measurement noise often overwhelms the underlying deformation signal. Second, we designed an exponentially convergent adaptive observer that reconstructs the internal state variable $\psi$ and estimates the frictional parameters $b$ and $d_c$ from the reconstructed friction signal. Third, we showed that iterative application of the observer across multiple fast-slip events allows progressive refinement of parameter estimates, achieving relative errors of approximately 15 \% for $(a-b)$ and 28 \% for $d_c$.

However, important limitations remain. The current implementation is restricted to the idealized spring-slider model, which cannot capture the spatial heterogeneity inherent to real faults. Moreover, the method estimates only a subset of RSF parameters ($a-b$ and $d_c$), requiring prior knowledge or assumptions about the direct-effect parameter $a$. Our observability analysis, however, indicates that during dynamic slip, all three parameters $a$, $b$, and $d_c$ are, in principle, observable. An important direction for future work is therefore to design an observer that can simultaneously estimate $a$, $b$, and $d_c$. Furthermore, future work should extend this framework by generalizing to spatially distributed fault models while maintaining the theoretical guarantees provided by the observer framework. Finally, our framework suggests a way to use observer theory as a tool for defining \emph{effective} frictional properties. In more realistic settings, one could regard a low-order spring--slider system as a reduced surrogate of a high-dimensional fault model, and use the observer to tune its RSF parameters so that its response matches either full-physics simulations or geophysical observations. In this interpretation, the estimated parameters represent spatially averaged frictional properties of a heterogeneous fault patch, providing a possible route to bridge detailed numerical models and model reduction descriptions without sacrificing the theoretical guarantees of the observer framework.

\bmsection*{Author contributions}
David Riley:  Software, Visualization, Writing – original draft.
 Diego Gutiérrez-Oribio:
Formal analysis, Software, Supervision, Writing – review \& editing. Ioannis Stefanou: Conceptualization, Supervision, Project administration, Funding acquisition, Writing – review \& editing.

\bmsection*{Acknowledgments}
The authors would like to acknowledge the European Research Council's (ERC) support under the European Union’s Horizon 2020 research and innovation program (Grant Agreement no. 101087771 INJECT).

\bmsection*{Funding}
This work was supported by the European Research Council's (ERC) support under the European Union’s Horizon 2020 research and innovation program (Grant Agreement no. 101087771 INJECT).

\bmsection*{Conflict of interest}

The authors declare no potential conflict of interest.

\bibliography{sample}

\bmsection*{Data Availability Statement}
No datasets were generated or analyzed during the
current study. Codes accompanying this manuscript will be made publicly available upon acceptance of this manuscript.
\appendix

\bmsection{Proof of the boundedness of the observer error}
\label{app:proof}
\vspace*{12pt}

We first introduce a convenient change of variable in the friction law as
\begin{equation}
  \bar{\mu}(t) := \mu\bigl(\lvert v(t)\rvert,\psi(t)\bigr) - a \ln\bigl(\lvert v(t)\rvert\bigr).
  \label{eq:mubar_def_proof}
\end{equation}
From the state evolution~\eqref{eq:state_evolution}, define
\begin{equation}
  \xi := \frac{1}{d_c}, 
  \qquad
  s(t) := v_* e^{-\psi(t)} - \lvert v(t)\rvert,
  \label{eq:theta_star_s_def}
\end{equation}
so that
\begin{equation}
\begin{split}
  \dot{\psi}(t) &= \xi s(t) \\
  \dot{\bar{\mu}}(t) &= b\,\xi\, s(t).
  \label{eq:mubar_dot_2}
\end{split}
\end{equation}

Define the estimation errors
\begin{equation}
  \tilde{\psi} := \psi - \hat{\psi},\qquad
  \tilde{b}   := b - \hat{b},\qquad
  \tilde{\xi} := \xi - \hat{\xi},
  \label{eq:param_errors}
\end{equation}
and the error signal for the observer
\begin{equation}
  e = \hat{\mu} - \bar{a} \ln\bigl(\lvert z_1\rvert\bigr) - \tilde{\mu}.
  \label{eq:e_def}
\end{equation}
Introduce the reconstruction errors of the friction and slip rate
\begin{equation}
\begin{split}
  \delta_\mu(t) &:= \hat{\mu}(t) - \mu\bigl(\lvert v(t)\rvert,\psi(t)\bigr), \\
  \delta_v(t) &:= z_1(t) - v(t),
  \label{eq:deltas_def_proof}
\end{split}
\end{equation}
and write
\begin{equation}
\begin{split}
  \hat{s}
    &= v_* e^{-\hat{\psi}} - \lvert z_1\rvert
    = s + \Delta_s,\\
  s &:= v_* e^{-\psi} - \lvert v\rvert, \\
  \Delta_s & := v_* \left(e^{-\hat{\psi}} - e^{-\psi} \right) - \lvert z_1\rvert + \lvert v\rvert
  \label{eq:s_splitting}
\end{split}
\end{equation}

Similarly, insert~\eqref{eq:mubar_def_proof} and~\eqref{eq:deltas_def_proof} into
\eqref{eq:e_def} to obtain
\begin{equation}
  e
  = \bar{\mu} - \tilde{\mu} + \Delta_e,
  \label{eq:e_splitting}
\end{equation}
where $\Delta_e = \delta_\mu + a\ln(\lvert v\rvert) - \bar{a}\ln(\lvert z_1\rvert)$.

Using Assumptions~\ref{A2}--\ref{A5} we can obtain
\begin{equation}
\begin{split}
  \lvert \Delta_s(t)\rvert &\le v_* L_\psi \lvert\tilde{\psi}(t)\rvert + \lvert\delta_v(t)\rvert , \\
  \lvert \Delta_e(t)\rvert &\le \lvert\delta_\mu(t)\rvert + a L_f \lvert\delta_v(t)\rvert + \tilde{a} \abs{\ln(\abs{z_1})},
  \label{eq:Delta_bounds}
\end{split}
\end{equation}
where $\tilde{a}=a-\bar{a}$.

Differentiate~\eqref{eq:e_splitting} and use~\eqref{eq:mubar_dot_2} and the first
line of~\eqref{eq:observer} to obtain
\begin{align*}
  \dot{e}
  &= \dot{\bar{\mu}} - \dot{\tilde{\mu}} + \dot{\Delta}_e
   = b\xi s - \hat{b}\hat{\xi}\hat{s} - k_1 e + \dot{\Delta}_e.
  \label{eq:e_dot_0}
\end{align*}
Expanding $\hat{b}\hat{\xi}=(b-\tilde{b})(\xi-\tilde{\xi})$ and using \eqref{eq:s_splitting} in the latter expression we obtain
\begin{equation}
  \dot{e}
  = -k_1 e
    + (b\tilde{\xi} + \xi\tilde{b} - \tilde{b}\tilde{\xi})\,s
    - \hat{b}\hat{\xi}\,\Delta_s
    + \dot{\Delta}_e.
  \label{eq:e_dot}
\end{equation}
Similarly, from~\eqref{eq:theta_star_s_def} and the second line of~\eqref{eq:observer},
\begin{align}
  \dot{\tilde{\psi}}
  &= \dot{\psi} - \dot{\hat{\psi}}
   = \xi s - \bigl(\hat{\xi}\hat{s} + k_2 e\bigr)
   = \tilde{\xi}s - k_2 e - \hat{\xi}\Delta_s.
  \label{eq:psi_tilde_dot}
\end{align}
Using the last two lines of~\eqref{eq:observer}, the parameter-error dynamics are
\begin{equation}
  \dot{\tilde{b}} = -k_3 e\,\hat{\xi}\hat{s},
  \qquad
  \dot{\tilde{\xi}} = -k_4 e\,\hat{b}\hat{s}.
  \label{eq:param_tilde_dot}
\end{equation}
Define the error vector $x := [e, \tilde{\psi}, \tilde{b}, \tilde{\xi}]^\top$. Following a Lyapunov analysis \cite{b:Khalil2002}, let us present the quadratic Lyapunov candidate
\begin{equation}
  V(x)
  := \frac{1}{2} e^2
     + \frac{1}{2}\,\tilde{\psi}^2
     + \frac{1}{2k_3}\,\tilde{b}^2
     + \frac{1}{2k_4}\,\tilde{\xi}^2.
  \label{eq:V_def_proof}
\end{equation}
Note that we could bound this function as
\begin{equation} 
\begin{split} 
\frac{m_1}{2} &\norm{x}^2 \leq V(x) \leq \frac{m_2}{2} \norm{x}^2, \\ 
m_1 &= \min \left\{1,\frac{1}{k_2},\frac{1}{k_3},\frac{1}{k_4} \right\}, \\ 
m_2 &= \max \left\{1,\frac{1}{k_2},\frac{1}{k_3},\frac{1}{k_4} \right\}. 
\end{split} 
\label{eq:Vbounds} 
\end{equation}

Its derivative along the trajectories~\eqref{eq:e_dot}--\eqref{eq:param_tilde_dot} is
\begin{align}
  \dot{V}
  &= e\,\dot{e}
   + \frac{1}{k_2}\tilde{\psi}\,\dot{\tilde{\psi}}
   + \frac{1}{k_3}\tilde{b}\,\dot{\tilde{b}}
   + \frac{1}{k_4}\tilde{\xi}\,\dot{\tilde{\xi}}
  \nonumber\\
  &= e\Bigl(
        -k_1 e
        + (b\tilde{\xi} + \xi\tilde{b} - \tilde{b}\tilde{\xi})\,s
        - \hat{b}\hat{\xi}\,\Delta_s
        + \dot{\Delta}_e
      \Bigr)
   \nonumber\\
  &\quad
   + \frac{1}{k_2}\tilde{\psi}\bigl(\tilde{\xi}s - k_2 e - \hat{\xi}\Delta_s\bigr)
   - \tilde{b}\,e\,\hat{\xi}\hat{s}
   - \tilde{\xi}\,e\,\hat{b}\hat{s}.
  \label{eq:Vdot_expand_proof}
\end{align}

Using \eqref{eq:s_splitting}, we rewrite the
last two terms as
\begin{align*}
  - \tilde{b} e \hat{\xi}\hat{s}
  - \tilde{\xi} e \hat{b}\hat{s}
  &= - \tilde{b} e \hat{\xi}(s+\Delta_s)
     - \tilde{\xi} e \hat{b}(s+\Delta_s) \\
  &= - \tilde{b} e \hat{\xi} s
                      - \tilde{\xi} e \hat{b} s
     - \tilde{b} e \hat{\xi}\Delta_s
                      - \tilde{\xi} e \hat{b}\Delta_s.
\end{align*}

Using Assumptions~\ref{A2}--\ref{A5}, the boundedness of $\psi,\hat{\psi}$ on
$\mathcal{K}$, and the bounds~\eqref{eq:Delta_bounds}, there exists $S_{\max}>0$
such that
\begin{equation}
  \lvert s(t)\rvert \le S_{\max},
  \qquad \forall t>0\ \text{with } \psi(t),\hat{\psi}(t)\in\mathcal{K}.
  \label{eq:s_bound}
\end{equation}
Moreover, on any local set
$
  \mathcal{X}_r := \left\{ x\in\mathbb{R}^4 \, : \, \|x\|\le r \, ; \, \psi,\hat{\psi}\in\mathcal{K}\right\},
$
we have
\begin{equation}
  |\tilde{b}(t)| \le r, \quad |\tilde{\xi}(t)| \le r,
  \quad
  \Rightarrow\quad
  |\tilde{b}(t)\tilde{\xi}(t)| \le r^2.
  \label{eq:local_r}
\end{equation}

We now collect separately the terms that contain only $s$ ($\dot{V}_{\mathrm{nom}}$) and those
containing $\Delta_s$ or $\dot{\Delta}_e$ ($\dot{V}_{\mathrm{pert}}$).

From~\eqref{eq:Vdot_expand_proof} and the cancellations described above, the
$s$-dependent (nominal) part of $\dot{V}$ is
\begin{equation}
  \dot{V}_{\mathrm{nom}}
  = -k_1 e^2 - \tilde{\psi}e
  + \frac{1}{k_2}\tilde{\psi}\tilde{\xi}s
  + e\,\tilde{b}\tilde{\xi}s.
  \label{eq:Vdot_nominal_part_final}
\end{equation}
Using Young’s inequality and the bounds
\eqref{eq:s_bound}--\eqref{eq:local_r}, we obtain
\begin{align}
  \dot{V}_{\mathrm{nom}}
  &\le -k_1 e^2
      + \Bigl(\tfrac{1}{2}e^2 + \tfrac{1}{2}\tilde{\psi}^2\Bigr)
      + \frac{S_{\max}}{2k_2}\tilde{\psi}^2 \nonumber \\
      &\quad  + \frac{S_{\max}}{2k_2}\tilde{\xi}^2
      + \frac{S_{\max} r^2}{2}e^2 \nonumber\\
  &\le -c_e e^2 + c_\psi \tilde{\psi}^2 + c_\xi \tilde{\xi}^2,
  \label{eq:Vdot_nominal_coeffs}
\end{align}
where the coefficients are explicitly
\begin{equation}
\begin{split}
  c_e   &:= k_1 - \frac{1}{2} - \frac{S_{\max} r^2}{2},\\[0.1em]
  c_\psi&:= \frac{1}{2} + \frac{S_{\max}}{2k_2},\\[0.1em]
  c_\xi &:= \frac{S_{\max}}{2k_2}.
\end{split}
\label{eq:coeffs_nominal}
\end{equation}

The term $\frac{1}{k_2}\tilde{\psi}(- \hat{\xi}\Delta_s)$ in \eqref{eq:Vdot_expand_proof} can be bounded using Assumption~\ref{A4}, \eqref{eq:param_errors} and \eqref{eq:s_splitting} as 
\begin{equation}
\begin{split}
    \frac{1}{k_2}\tilde{\psi}(- \hat{\xi}\Delta_s) &= -\frac{1}{k_2}\tilde{\psi}\xi\Delta_s + \frac{1}{k_2}\tilde{\psi}\tilde{\xi}\Delta_s \\
    &\leq -\frac{v_* L_\psi}{k_2}\xi \tilde{\psi}^2 + \frac{1}{k_2}\tilde{\psi}\tilde{\xi}\Delta_s.
\end{split}
\label{eq:k_2_term}
\end{equation}

Substituting~\eqref{eq:k_2_term} into
$\dot{V}_{\mathrm{nom}}$ we get
\begin{align}
  \dot{V}_{\mathrm{nom}}
  &\le -c_e e^2 - c_\psi^* \tilde{\psi}^2 + c_\xi \tilde{\xi}^2,
  \label{eq:Vdot_nom_final}
\end{align}
with
\begin{equation}
  c_\psi^*:= \frac{v_* L_\psi}{k_2}\,\xi
            - \frac{1}{2} - \frac{S_{\max}}{2k_2}.
\label{eq:coeffs_nominal_updated}
\end{equation}
Therefore, for any fixed $r>0$ we can choose the gains $k_1,k_2$ as \eqref{eq:gains_obs}, such that
\begin{equation}
  \dot{V}_{\mathrm{nom}}
  \le -\alpha_0\bigl(e^2 + \tilde{\psi}^2\bigr)
     + c_\xi \tilde{\xi}^2,
  \label{eq:Vdot_nom_compact}
\end{equation}
with $\alpha_0 := \min\{c_e,c_\psi\} > 0$.

\medskip
All remaining terms in~\eqref{eq:Vdot_expand_proof} involve $\Delta_s$ or
$\dot{\Delta}_e$ and can be grouped as
\[
  \dot{V}_{\mathrm{pert}}
  =
   e\bigl(- \hat{b}\hat{\xi}\,\Delta_s + \dot{\Delta}_e\bigr)
   + \frac{1}{k_2}\tilde{\psi}\tilde{\xi}\Delta_s
   - \tilde{b} e \hat{\xi}\Delta_s
   - \tilde{\xi} e \hat{b}\Delta_s.
\]
Using the bounds~\eqref{eq:Delta_bounds}, the local boundedness of
$\hat{b},\hat{\xi}$, and the constraint $\|x\|\le r$, there exist constants
$c_1,c_2>0$ such that
\begin{equation}
  |\dot{V}_{\mathrm{pert}}|
  \le c_1\|x(t)\|^2 + c_2\|x(t)\|\,\|\delta(t)\|.
  \label{eq:Vdot_pert_bound_final}
\end{equation}
Combining \eqref{eq:Vdot_nom_compact} and \eqref{eq:Vdot_pert_bound_final} yields
\begin{equation}
  \dot{V}
  \le -\alpha_0\bigl(e^2 + \tilde{\psi}^2\bigr)
     + (c_\xi + c_1)\|x(t)\|^2
     + c_2\|x(t)\|\,\|\delta(t)\|.
  \label{eq:Vdot_before_alpha}
\end{equation}
Moreover, by local boundedness of the parameter errors on $\mathcal{X}_r$ and
the structure of $V$, there exists a constant $\beta\in(0,1]$ such that
\begin{equation}
  e^2 + \tilde{\psi}^2 \;\ge\; \beta\,\|x(t)\|^2,
  \qquad \forall\,x\in\mathcal{X}_r.
  \label{eq:e_psi_lower_bound}
\end{equation}
Using~\eqref{eq:e_psi_lower_bound} in~\eqref{eq:Vdot_before_alpha}, we obtain, for all
$x\in\mathcal{X}_r$,
\begin{equation}
\begin{split}
  \dot{V}
  &\le -\alpha_0 \beta \|x(t)\|^2
     + (c_\xi + c_1)\|x(t)\|^2
     + c_2\|x(t)\|\,\|\delta(t)\| \\
  &\le -\bigl(\alpha_0\beta - c_\xi - c_1\bigr)\|x(t)\|^2
     + c_2\|x(t)\|\,\|\delta(t)\|.
\end{split}
\end{equation}
Hence, there is a sufficiently big $\alpha_0>0$ such that, for all $x\in\mathcal{X}_r$, it yields
\begin{equation}
  \dot{V}
  \le -\alpha \|x(t)\|^2 + c_2\|x(t)\|\,\|\delta(t)\|.
  \label{eq:Vdot_ISS_step2}
\end{equation}
Using Young’s inequality,
$
  c_2\|x\|\,\|\delta(t)\|
  \le \frac{\alpha}{2}\|x\|^2 + \frac{c_2^2}{2\alpha}\|\delta(t)\|^2,
$
and the bounds \eqref{eq:Vbounds}, we obtain the eISS inequality
\begin{equation}
  \dot{V}
  \le - \frac{\alpha}{m_2} V + \frac{c_2^2}{2\alpha} \|\delta(t)\|^2.
  \label{eq:Vdot_ISS_final}
\end{equation}
From the differentiator and friction bounds~\eqref{eq:dif_accu}--\eqref{eq:mu_accu},
there exists $D>0$ such that
\begin{equation}
  \sup_{t\geq 0} \|\delta(t)\|
  \leq D
  := \frac{L \rho^2}{h}
       \Bigl[m \gamma_2 + 2 h \gamma_1 \rho + k \gamma_0 \rho^2 \Bigr].
  \label{eq:delta_sup_bound_final}
\end{equation}
Applying the comparison lemma to~\eqref{eq:Vdot_ISS_final} yields, for all $t\geq 0$,
\begin{equation}
  V(t)
  \leq e^{-\frac{\alpha}{m_2} t} V(0)
   + \frac{c_2^2 m_2}{2 \alpha^2}
     D^2.
  \label{eq:V_bound_final}
\end{equation}
Using~\eqref{eq:Vbounds}, we finally obtain
\begin{equation}
  \|x(t)\|
  \leq \sqrt{\frac{m_2}{m_1}} e^{-\frac{\alpha}{2 m_2} t} \|x(0)\|
      + \frac{c_2}{\alpha}\sqrt{\frac{m_2}{2 m_1}}\,D,
  \label{eq:psi_accu_final_clean}
\end{equation}
for all $t\geq 0$ such that $x\in\mathcal{X}_r$. This provides local exponential convergence of the error vector $x = [e,\,\tilde{\psi},\,\tilde{b},\,\tilde{\xi}]^\top$ to an ultimate bound which depends explicitly on the differentiator accuracy \eqref{eq:dif_accu} and the friction reconstruction accuracy~\eqref{eq:mu_accu}. 
\hfill $\qed$




\end{document}